\documentclass[a4pape,citesort]{PoS}
\usepackage{caption}
\usepackage{subcaption}
\usepackage{graphicx}
\usepackage{citesort}
\usepackage{bold-extra}
\usepackage{afterpage}
\usepackage[backend=bibtex,style=numeric,sorting=none,doi=false]{biblatex}
\usepackage{lipsum,lineno}
\RequirePackage{amsmath}
\RequirePackage{xspace}
\newcommand{ \xfitter} {\textsc{xFitter}\xspace}
\newcommand{\xbj}{\ensuremath{x_{\text{Bj}}}\xspace}

\usepackage{wrapfig}
\usepackage{lipsum,lineno}
\usepackage{microtype}  %%% HELP WITH HYPHENATION

\makeatletter 
\renewcommand\speaker[1]{\if@speaker\global\@dblspeaktrue\fi
			\global\@speakertrue
			\global\setbox\@firstaubox
			\hbox{{\let\thanks\@gobble
				\let\footnote\@gobble\small
				\rm  The nCTEQ Collaboration}}%
			#1\thanks{Speaker.}\
			}%
\makeatother

\title{Recent QCD results from the \xfitter project:
\hspace{\textwidth} 
{\it Probing the strange content of the proton with charm production in charged current at LHeC}
}

\ShortTitle{xFitter: Charm Production at the LHeC }

\def\thanksref#1{\rlap,${}^{#1}$}
\def\inst#1{\hangafter=1\hangindent=15pt\relax ${}^{#1}$}

\author{
The \xfitter Developers' Team:\thanks{%
We acknowledge the hospitality of CERN, DESY, and Fermilab where a
portion of this work was performed.
This work was also partially supported by the U.S.\ Department of
Energy under Grant No.\ DE-SC0010129. 
We are grateful to the DESY IT department for their support of the
\xfitter developers.
}
\quad
Hamed~Abdolmaleki\thanksref{a}
Valerio~Bertone\thanksref{b}
Daniel~Britzger\thanksref{c}
Stefano~Camarda\thanksref{d}
Amanda~Cooper-Sarkar\thanksref{e}
Achim~Geiser\thanksref{f}
Francesco~Giuli\thanksref{g}
Alexander~Glazov\thanksref{f}
Agnieszka~Luszczak\thanksref{h}
Ivan~Novikov\thanksref{i}
Fred~Olness\thanksref{j}\speaker{}
Andrey Sapronov\thanksref{i}
Oleksandr~Zenaiev\thanksref{k}
\\
\inst{a} Faculty of Physics, Semnan University, 35131-19111 Semnan,  Iran   \\
\inst{b} Dipartimento di Fisica, Universit\`a di Pavia and INFN, Sezione di Pavia Via Bassi 6, I-27100 Pavia, Italy   \\
\inst{c} Max-Planck-Institut f\"ur Physik, F\"ohringer Ring 6, D-80805 M\"unchen, Germany   \\
\inst{d} CERN, CH-1211 Geneva 23, Switzerland   \\
\inst{e} Particle Physics, Denys Wilkinson Bdg, Keble Road,  University of Oxford, OX1 3RH Oxford,~UK   \\
\inst{f} Deutsches Elektronen-Synchrotron (DESY), Notkestrasse 85,  D-22607 Hamburg, Germany   \\
\inst{g} University of Rome Tor Vergata and INFN, Sezione di Roma 2, Via  della Ricerca Scientifica 1, 00133 Rome, Italy   \\
\inst{h} T. Kosciuszko Cracow University of Technology, PL-30-084, Cracow, Poland   \\
\inst{i} Joint Institute for Nuclear Research, Joliot-Curie 6, Dubna, Moscow region, Russia, 141980   \\
\inst{j} SMU Physics, Box 0175 Dallas, TX 75275-0175, USA   \\
\inst{k} Hamburg University, II. Institute for Theoretical Physics,   Luruper Chaussee 149, D-22761 Hamburg, Germany   \\
}

\abstract{%
We investigate charm production in charged-current deep-inelastic
scattering (DIS) using the  \xfitter program.
 \xfitter is an open-source software framework for the determination
of PDFs and the analysis of QCD physics, and has been used for a
variety of LHC studies.
The study of charged current DIS charm production provides an
important perspective on the strange quark PDF, $s(x)$. We make use of
the \xfitter tools to study the present $s(x)$ constraints, and then
use LHeC pseudodata to infer how these might improve.
Furthermore, as \xfitter implements both Fixed Flavor and Variable
Flavor number schemes, we can examine the impact of these different
theoretical choices; this highlights some interesting aspects of
multi-scale calculations.
This study provides a practical illustration of the many features of
xFitter.
}

\FullConference{XXVII International Workshop on Deep-Inelastic Scattering and Related Subjects - DIS2019\\
		8-12 April, 2019\\
		Torino, Italy}

\begin{document}
%\linenumbers

%%%%%%%%%%%%%%%%%%%%%%%%%%%%%%%%%%%%%%%%%%%%%%%%%%%%%%%%%%%%
\def\figCompQ{
% FIG 1
\begin{figure}[t]
\null
\vspace{-0.4cm}
  \centering
    \includegraphics[width=0.80\textwidth]{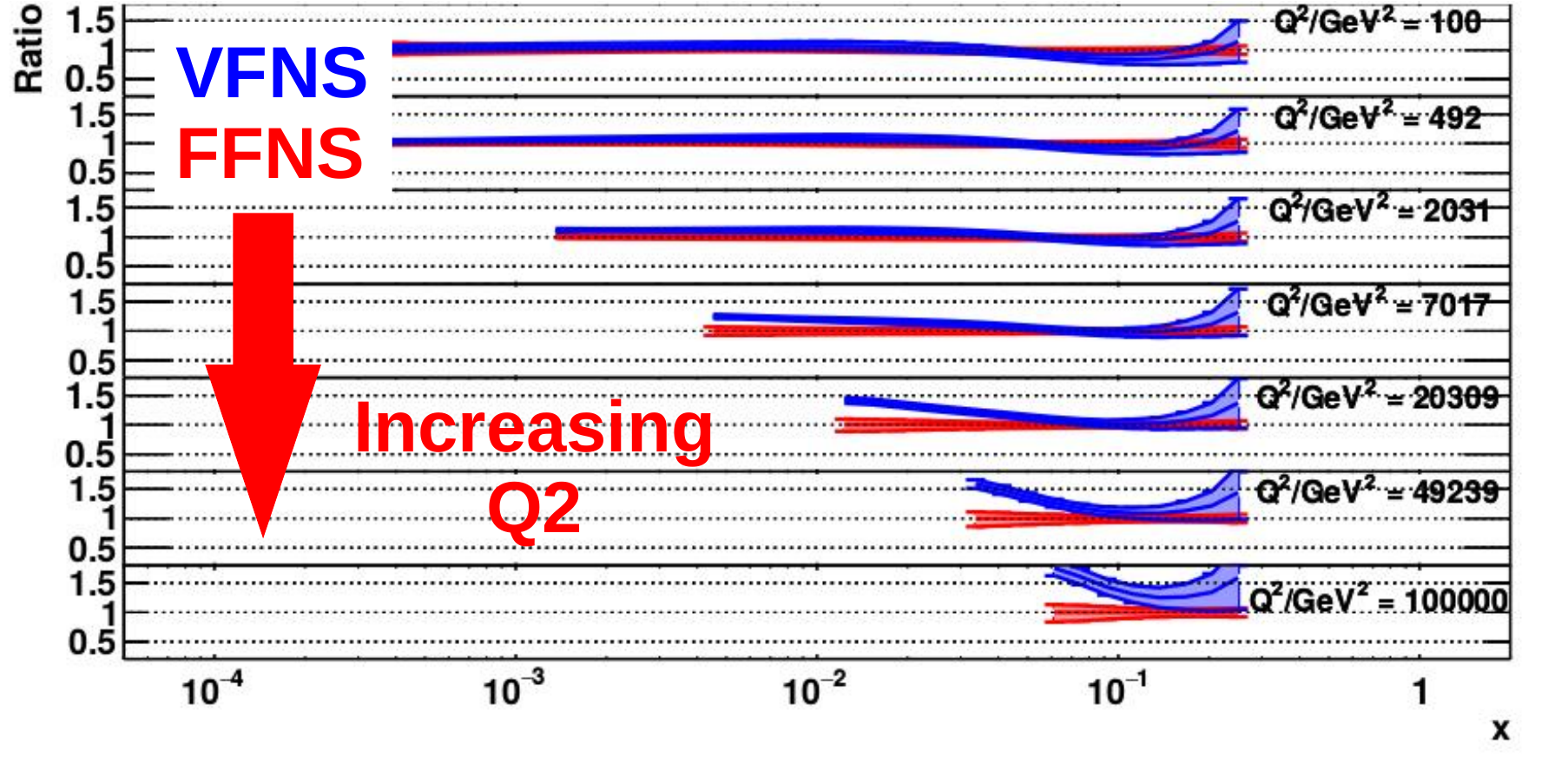}
    \includegraphics[width=0.80\textwidth]{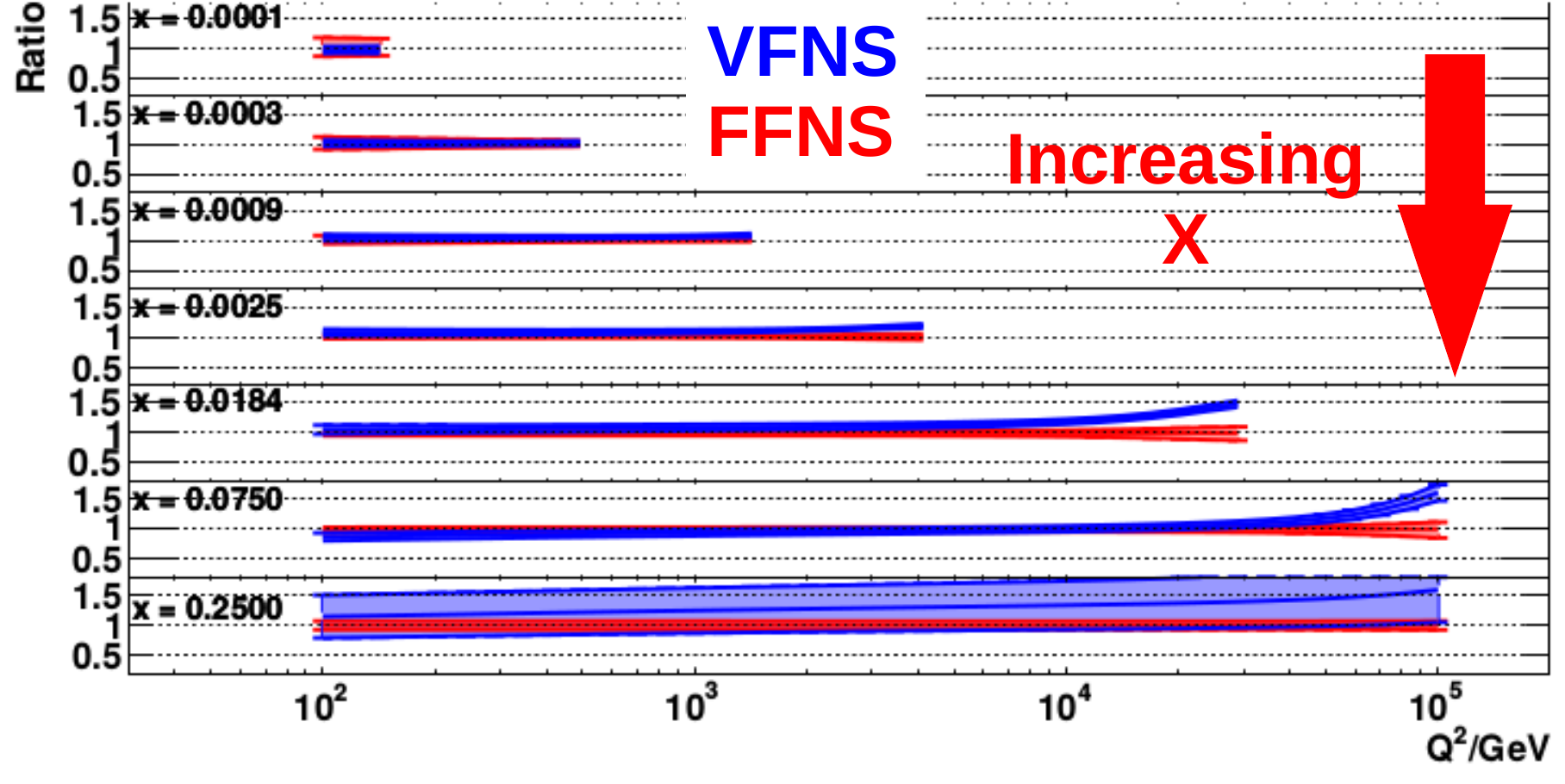}
\vspace{-0.3cm}
    \caption{
Comparison of the theoretical predictions (ratio) with  uncertainties for
CC charm production at the LHeC vs.
$x_{bj}$~(top) and  $Q^2$~(bottom)
as calculated in the FFNS (FFNS~A) and VFNS (FONLL-B) context.
We observe increasing differences at large $Q^2$ and large $x_{bj}$ values.\cite{Abdolmaleki:2019acd}
\label{fig:compQ2}
}
\vspace{-0.5cm}
\end{figure}
}
%%%%%%%%%%%%%%%%%%%%%%%%%%%%%%%%%%%%%%%%%%%%%%%%%%%%%%%%%%%%

%%%%%%%%%%%%%%%%%%%%%%%%%%%%%%%%%%%%%%%%%%%%%%%%%%%%%%%%%%%%
\def\figFlav{
% FIG 2
\begin{figure}[t]
\null
\vspace{-0.3cm}
  \centering
    \includegraphics[width=0.95\textwidth]{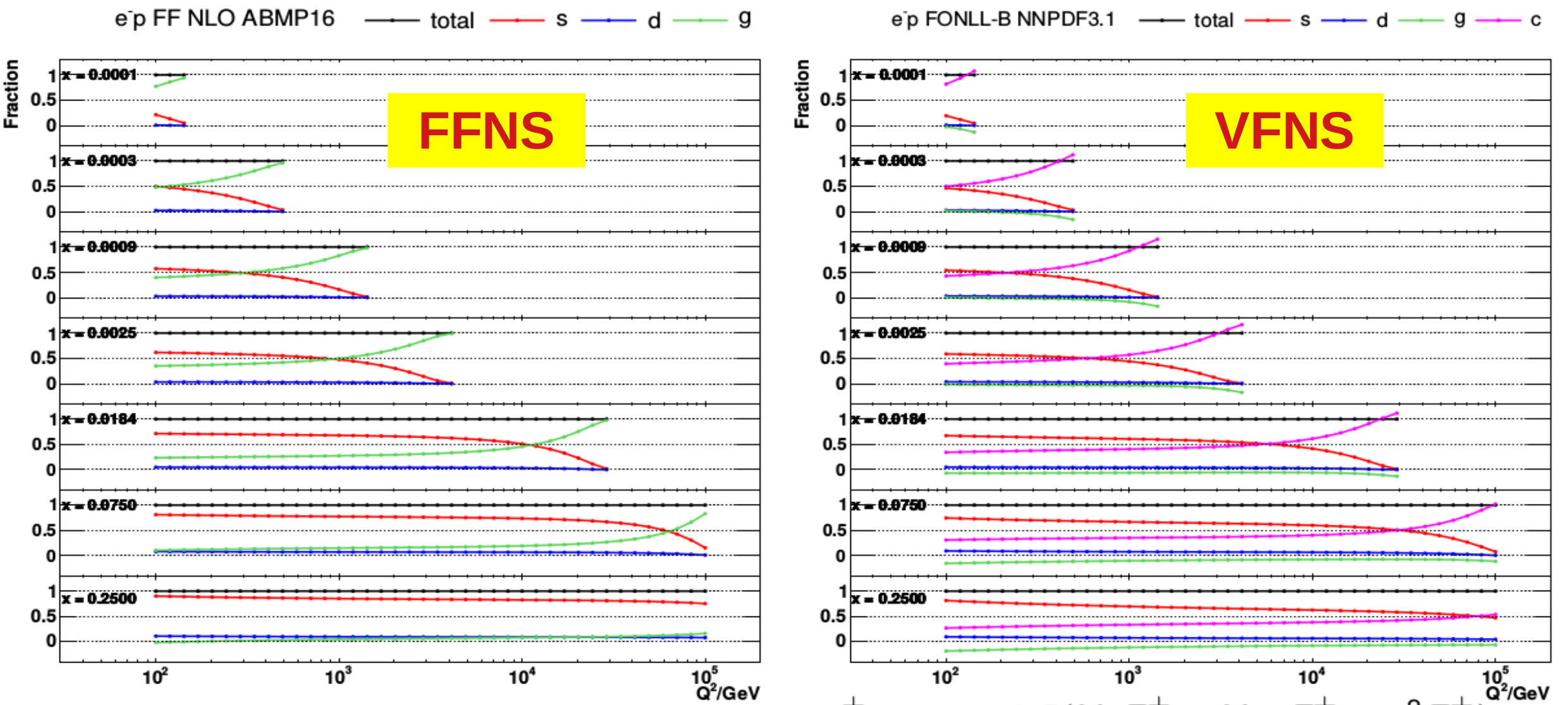}
    \caption{
The partonic subprocesses for charm CC production cross sections for
 FFNS (FFNS~A) and VFNS (FONLL-B)  as a function of $Q^2$ 
for different values of $x_{bj}$, Ref.~\cite{Abdolmaleki:2019acd}.
\label{fig:Flavor}
}
\vspace{-0.5cm}
\end{figure}
}
%%%%%%%%%%%%%%%%%%%%%%%%%%%%%%%%%%%%%%%%%%%%%%%%%%%%%%%%%%%%

%FIG 3
%%%%%%%%%%%%%%%%%%%%%%%%%%%%%%%%%%%%%%%
\def\figBandS{
\begin{figure}[t]
%\begin{wrapfigure}{R}{8cm} %%%%%%%%%%%%%%%%%%%%%
\centering{} 
  \begin{subfigure}[t]{0.50\textwidth}
    \centering
    \includegraphics[width=0.99\textwidth]{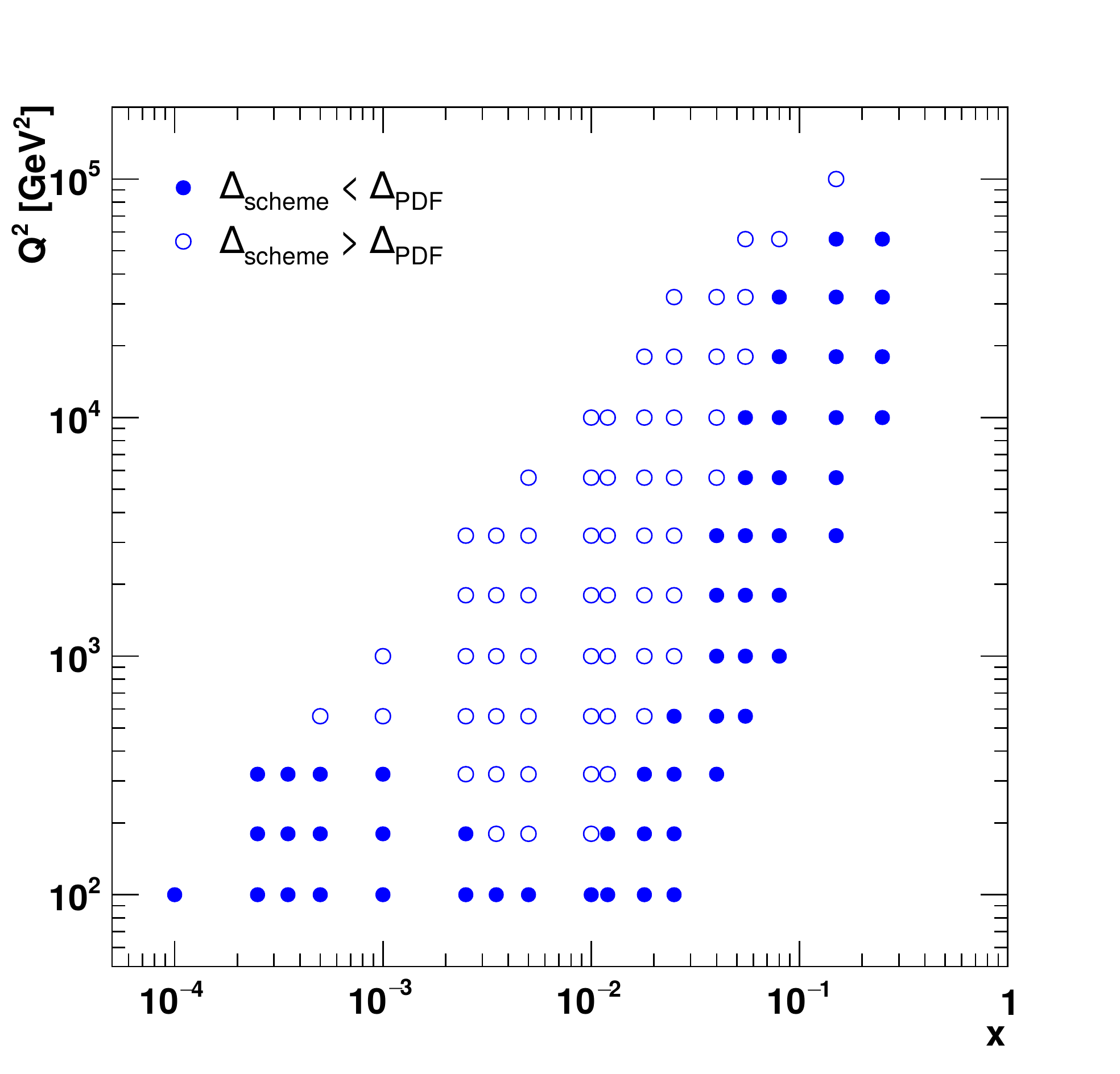}
    \caption{The full ($\Delta_{\rm scheme} < \Delta_{\rm PDF}$, $\Delta_{\rm scheme} > \Delta_{\rm PDF}$)
    and restricted ($\Delta_{\rm scheme} < \Delta_{\rm PDF}$) sets of data points which are used for PDF profiling.}
\label{fig:pseudo}
 \end{subfigure}
\hfil
  \begin{subfigure}[t]{0.45\textwidth}
   \centering
    \includegraphics[width=0.99\textwidth]{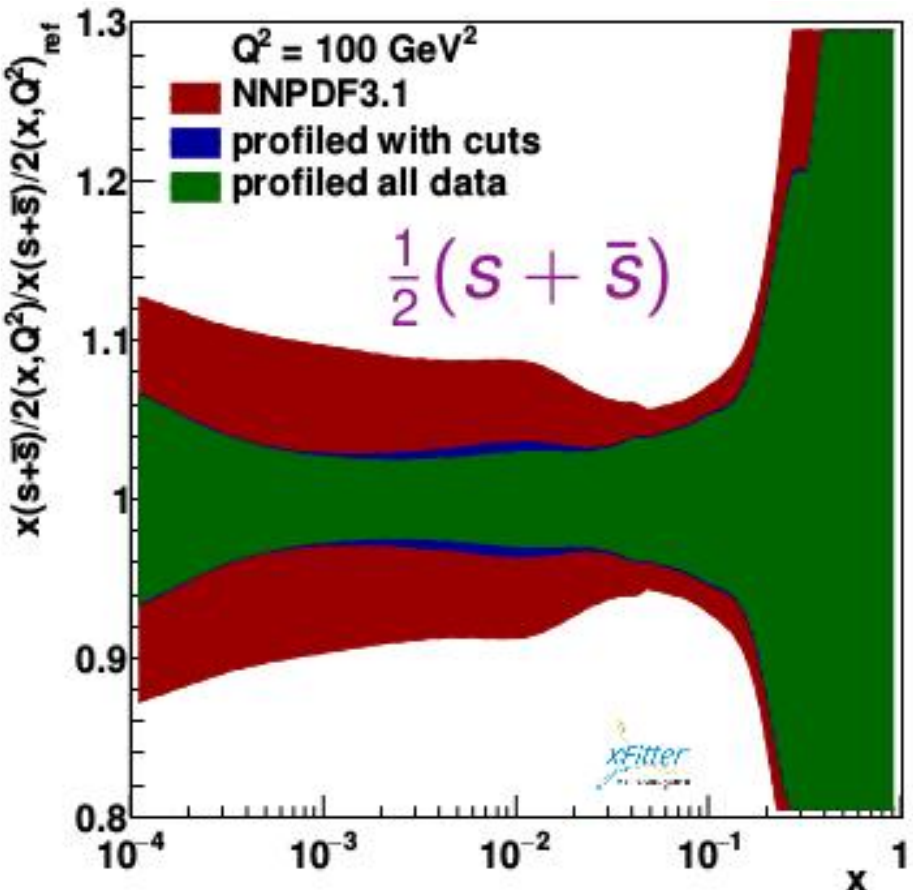}
    \caption{The relative strange  quark PDF uncertainties at
       $Q^2$=100~GeV${}^2$ of the original and
       profiled NNPDF3.1 PDF set.}
     \label{fig:bandS}
    \end{subfigure}
%\vspace{-15pt}
\caption{LHeC Constraints on the Strange PDF, {\it c.f.}, Ref.~\cite{Abdolmaleki:2019acd}.}
%\vspace{-20pt}
%\end{wrapfigure}
\label{fig:three}
\end{figure}
}
%%%%%%%%%%%%%%%%%%%%%%%%%%%%%%%%%%%%%%%%%%%%

%%%%%%%%%%%%%%%%%%%%%%%%%%%%%%%%%%%%%%%%%%%%%%%%%%%%%%%%%%%%
\section{ \xfitter Overview}

The Parton Distribution Functions (PDFs) are the essential components
that allow us to make theoretical predictions for experimental
measurements of  hadrons.
The precision of the PDF analysis has advanced tremendously  in recent years, 
and these studies are now performed with very high  precision  at NLO and  NNLO in perturbation theory.
The \xfitter project\footnote{%
xFitter can be downloaded from {\tt www.xFitter.org}.  An overview of
the program can be found in Ref.~\cite{Alekhin:2014irh}.
\\
A more extensive report of these features can be found in Ref.~\cite{Abdolmaleki:2019acd}.
Also see the recent study of Ref.~\cite{AbdulKhalek:2019mps}.  
} 
is an open source QCD fit framework 
that can perform PDF fits, assess the impact of new data, compare existing PDF sets, 
and perform a variety of other tasks~\cite{Alekhin:2014irh}.
The modular structure of \xfitter allows for interfaces to 
a variety of external programs including: 
 QCDNUM~\cite{Botje:2010ay},
 APFEL~\cite{Bertone:2013vaa},
 LHAPDF~\cite{Buckley:2014ana},
 APPLGRID~\cite{Carli:2010rw},
 APFELGRID~\cite{Bertone:2016lga},
 FastNLO~\cite{Britzger:2012bs}  
and
 HATHOR~\cite{Aliev:2010zk}. 
%
%% \color{red}
%%  A schematic of the modular structure is illustrated in Fig.~\ref{fig:flow}.
%% \color{black}
%

An overview of the recent \xfitter updates and available tutorials 
is available in Ref.~\cite{Bertone:2017tig}. 
In this short report we will focus on the charged current (CC) production of charm
and the impact on the underlying strange quark
PDF~\cite{Abdolmaleki:2019acd}.
%  
%\footnote{A more extensive report of these features can be found in Ref.~\cite{Abdolmaleki:2019acd}.}

%%%%%%%%%%%%%%%%%%%%%%%%%%%%%%%%%%%%%%%%%%%%%%%%%%
%\newpage
\null \vspace{-0.8cm}
\section{Charged Current (CC) Charm Production and the LHeC}
%\vspace{-0.3cm}

%%%%%%%%%%%%%
%\newpage

In this study, we will examine the charged current (CC) production of charm
to provide insight on the underlying strange quark PDF.
The strange quark PDF has been extensively investigated in a number of
experiments including
fixed-target neutrino/antineutrino-nucleon DIS experiments,
and the associated production of a $W$~boson
with a charm-jet final state at the LHC.
Despite these measurements,
$s(x)$ still has a sizable uncertainty; in the future it is essential  to reduce this
uncertainty as we strive to make increasingly
precise tests of the SM and search for what might lie beyond.

The proposed Large Hadron Electron Collider (LHeC) facility can
provide high statistics measurements of electrons on both
protons and nuclei across a broad kinematic range to address many of
these outstanding questions~\cite{AbelleiraFernandez:2012cc}.
For example,  a 7~TeV proton beam on a 60~GeV
electron beam could provide $\sqrt{s}\sim 1.3$~TeV.
Compared to HERA, the LHeC extends the covered kinematic range by an
order of magnitude in both \xbj and $Q^2$ with a nominal design
luminosity of $10^{33} cm^{-2} s^{-1}$.

Specifically, the process we'll consider is
$W s\to c$ at leading-order (LO);  higher-order corrections
include  $W g \to c \bar{c}$ and $W c \to c \bar{s}$. 
This process touches on a number of interesting QCD aspects.
It is a multi scale problem as it involves  the boson mass $M_W$ and $Q^2$ scale
in addition to the quark masses $\{m_s,m_c\}$.
Additionally, because  we have possible contributions from a charm PDF at NLO,
we can choose to compute this in either the Variable Flavor Number Scheme (VFNS)
or a Fixed Flavor Number Scheme (FFNS)
As  \xfitter has both VFNS and FFNS implemented, we find it useful to compare these
two perspectives.

\figCompQ
%%%%%%%%%%%%%%%%%%%%%%%%%%%
\section{The VFNS and FFNS}
%\vspace{-0.3cm}

In Fig.~\ref{fig:compQ2} we display the ratio for  NLO CC charm production 
computed in the VFNS (using the FONLL-B calculation with the NPDF3.1 NLO PDF set),
and in the FFNS (using the FFNS~A calculation with the ABMP16 NLO PDF set).
Examining Fig.~\ref{fig:compQ2}-a) we observe that the VFNS and FFNS yield comparable
results throughout the $x$ range for lower $Q^2$ values, but differ for increasing $Q^2$.
Correspondingly, in Fig.~\ref{fig:compQ2}-b) the differences increase for the larger $x$ values~\cite{Abdolmaleki:2019acd}.

{\bf $Q^2$ Dependence:}
To better understand the relative behavior of the  FFNS
and VFNS calculations, we recall that when the scale $Q\sim \mu$ is below
the charm PDF matching scale $\mu_c$ (typically taken to be equal
to $m_c$~\cite{Bertone:2017ehk}) the charm PDFs vanish and the FFNS and
VFNS reduce to the same result.\footnote{Note,we note the PDF uncertainties are taken from the separate underlying PDF sets
with their distinctive methodologies  and cannot be directly compared.}
For increasing $Q$ scales, 
the VFNS resums the
$\alpha_S \ln(Q^2/m_c^2)$
 contributions via the
DGLAP evolution equations and the FFNS and VFNS will
slowly diverge logarithmically. This behavior is observed in
Fig.~\ref{fig:compQ2} and is consistent with the characteristics demonstrated
in Ref.~\cite{Kusina:2013slm}.
Thus, we have identified the
source of the scheme differences at large $Q^2$.

{\bf $x$ Dependence:}
The source of the scheme differences at large $x$ is a bit
more intricate. The VFNS uses the DGLAP equations to
resum  higher-order logarithms of the form $\alpha_S \ln(Q^2/m_c^2)$,
and these are balanced with the counter-terms of the NLO contributions.
Thus, the net contribution of the VFNS over the FFNS depends on
the particular value of $x$, and we find  (c.f.,~Fig.~11 of Ref.~\cite{Kusina:2013slm}.)
that these terms increase for larger $x$ values.
Thus, we have identified the source
of the scheme differences at large $x$.

%% {\bf PDF Uncertainty:}
%% %
%% Finally, we note the PDF uncertainties are taken from the separate underlying PDF sets
%% with their distincitive methodologies 
%% and cannot be directly compared.
%% %
%% A more direct comparison is enabled by by using a set of PDFs matched
%% at $Q_0<m_c$ (as in Fig.~4 of Ref.~\cite{Kusina:2013slm}) or
%% by  examining the $\mu$ scale dependence alone (as in Figs.~4 and~5 of Ref.~\cite{Abdolmaleki:2019acd}).

\figFlav
%%%%%%%%%%%%%%%%%%%%%%%%%%%
\section{Flavor Decomposition}
%\vspace{-0.3cm}

In Fig.~\ref{fig:Flavor} 
we display the separate partonic contributions to the total cross section
as a function of $\{x,Q^2\}$ in both the FFNS and VFNS. 
We find it very interesting that the charm contribution in the VFNS (magenta line) is remarkably
similar to the gluon contribution in the FFNS (green line).
In the FFNS decomposition, we note that the charm component is not present by definition as
the charm PDF is not included in this framework; charm can only enter here via an external
gluon splitting process $g\to c \bar{c}$.

In contrast, in the VFNS the charm component is present,
and mainly produced by $g\to c \bar{c}$ collinear splitting through
DGLAP evolution. The fundamental underlying process is
(and has to be) the same in both the FFNS and VFNS, but the
factorization boundary between PDFs and hard scattering
cross section (determined by the scale $\mu$ and the
scheme choice) is different.
This behavior underscores
the fact that the renormalization scale $\mu$ is simply ``shuffling''
contributions among the separate sub-pieces, but the
total physical cross section remains positive and stable,
{\it cf.}~Refs.~\cite{Abdolmaleki:2019acd,Kusina:2013slm}.
 This is a triumph of the QCD theory.

Next, turning our attention to the strange PDF contribution,
we observe  that the FFNS and VFNS  predictions behave
qualitatively very similar as functions of the kinematic variables. 
Specifically, the strange fraction increases for 
$x_{Bj}$ and decreases for $Q^2$ and $y$. In particular, at high $y$ the
strange PDF contribution drops to zero in favor of the gluon
or charm quark PDFs.  In these phase-space regions, the
dominant contributions to the cross section are proportional
to the gluon PDF in the FFNS or to the charm-quark PDFs
in the VFNS.

Finally, we add that the  \xfitter 2.0.0 program links to the APFEL code~\cite{Bertone:2013vaa}
which has implemented generalized matching conditions that enable
the switch from NF to NF +1 active flavors at an arbitrary matching scale $\mu_m$. This enables us to
generalize the transition between a FFNS and a VFNS and essentially vary continuously between
the two schemes; in this sense the matching scale  $\mu_m$ allows us to unify the FFNS and VFNS in a
common framework~\cite{Bertone:2017ehk}.

\figBandS
%%%%%%%%%%%%%%%%%%%%%%%%%%%
\section{LHeC Constraints on the Strange PDF}
%\vspace{-0.3cm}

Having outlined the theoretical ingredients we now
assess the ability of the LHeC to constrain the PDFs. 
We use pseudodata for differential CC charm production cross sections
in $Q^2$ and \xbj  corresponding to an integrated
luminosity of $100{\rm~fb}^{-1}$ and
polarization $P=-0.8$.
The charm-mass reference value in the
$\overline{\mbox{MS}}$ scheme is set to $m_c(m_c) = 1.27$ GeV and
$\alpha_s$ is set to the value used for the corresponding PDF
extraction. The renormalization and factorization scales are chosen to
be $\mu_\mathrm{r}^2 = \mu_\mathrm{f}^2 = Q^2$.
Details are provided in Ref.~\cite{Abdolmaleki:2019acd}.

We have observed in the previous discussion that the VFNS and FFNS can differ
across the $\{x,Q^2\}$ kinematic plane. This is illustrated in Fig.~\ref{fig:pseudo}
where the open data points indicate where the difference between the  VFNS and FFNS
is larger than the PDF uncertainty, $\Delta_{\rm scheme} > \Delta_{\rm PDF}$.
To gauge the impact of this heavy flavor scheme choice, we will therefore 
perform the profiling study of the PDFs
with i)~the full set of LHeC pseudodata, and
ii)~the restricted set  (with cuts)  where the difference between the  VFNS and FFNS
is smaller than the PDF uncertainty, $\Delta_{\rm scheme} < \Delta_{\rm PDF}$ (solid data points.).
This latter profiled PDF will provide a conservative estimate that is independent of
the particular VFNS or FFNS, 

In Fig.~\ref{fig:bandS} we display the strange PDF uncertainty for the
the original PDF set, the profiled PDF set with the all data, and with the PDF with the restricted data set with cuts. 
For both LHeC data sets, the reduction in  $s(x)$ is dramatic in the intermediate to low $x$ region;
this is encouraging as it demonstrates the LHeC will have significant impact on the strange PDF
across a broad kinematic region. 
Additionally, we note that the difference between the full data and the restricted data is minimal;
hence, any uncertainty arising  from the VFNS/FFNS choice has a negligible impact on the final PDF.

%%%%%%%%%%%%%%%%%%%%%%%%%%%%%%%%%%%%%%%%%%%%%%%%%%
%%%%%
\newpage
\null \vspace{-1.5cm}
\section{Conclusion}
 \vspace{-0.4cm}

The \xfitter 2.0.0 program is a versatile, flexible, modular, and
comprehensive tool that can facilitate analyses of the experimental
data and theoretical calculations.
In this study we have examined the ability of 
CC charm production data at the LHeC to constraint the PDF uncertainty.
This project demonstrates a number of the  \xfitter features including
the ability to compute in both the VFNS and the FFNS. 
This feature, in addition to the variable heavy flavor matching scale $\mu_m$,
allow us to generalize the transition between a VFNS and a FFNS, and provides a
theoretical ``laboratory'' which can quantitatively
explore aspects of the QCD theory. 
We have decomposed  the separate flavor components
contributing to the process, and
studied the  effects of the  VFNS and FFNS
across the  $\{x,Q^2\}$ kinematic plane.

Finally, we use the profiling ability of  \xfitter to
investigate the impact of the LHeC pseudodata and find
this can dramatically improve the PDF uncertainty
{\em independent} of the underlying heavy flavor scheme used in the calculation.
This study not only provides new insights into the
intricacies of QCD, but also has practical advantages for PDF fits
and future facilities.

%%%%%%%%%%%%%%%%%%%%%%%%%%%%%%%%%%%%%%%%%%%%%%%%%%
%\section{Acknowledgment}
%%%%%%%%%%%%%%%%%%%%%%%%%%%%%%%%%%%%%%%%%%
%\bibliography{refs}{}
%\bibliography{refs,extra}{}
%\bibliographystyle{plain}
%\bibliographystyle{plainnat}
%\bibliographystyle{unsrt}
%\bibliographystyle{hep}
%\bibliographystyle{hep}
%\bibliographystyle{hep}
%\bibliographystyle{h-physrev}
%\bibliographystyle{agsm}
%\bibliographystyle{abbrv}
%\bibliographystyle{habbrv}

\printbibliography 
%\input{dis19.bbl}

%\input{main_final.bbl}
%\printbibliography
%%%%%%%%%%%%%%%%%%%%%%%%%%%%%%%%%%%%%%%%%%
\end{document}